\documentclass[aps,prc,reprint,nofootinbib,superscriptaddress]{revtex4-1}
\usepackage{epsfig}
\usepackage{multirow}
\usepackage{amsmath}
\usepackage{amssymb}
\begin{document}


\title{Measurement of cross sections and astrophysical S-factors for \textit{d(p,$\gamma$)$^3$He} reaction at astrophysically relevant energies using a large volume LaBr$_3$:Ce detector}


\author{M. Dhibar}
\affiliation{Department of Physics, Indian Institute of Technology - Roorkee, 247667, INDIA}
\author{I. Mazumdar}
\email{indra@tifr.res.in}
\affiliation{Deptartment of Nuclear and Atomic Physics, 
Tata Institute of Fundamental Research, Mumbai - 400005, INDIA}
\author{G. Anil Kumar}
\affiliation{Department of Physics, Indian Institute of Technology - Roorkee, 247667, INDIA}
\author{A. K. Rhine Kumar}
\affiliation{Deptartment of Nuclear and Atomic Physics, 
Tata Institute of Fundamental Research, Mumbai - 400005, INDIA}
\author{S. M. Patel}
\affiliation{Deptartment of Nuclear and Atomic Physics, 
Tata Institute of Fundamental Research, Mumbai - 400005, INDIA}
\author{P. B. Chavan}
\affiliation{Deptartment of Nuclear and Atomic Physics, 
Tata Institute of Fundamental Research, Mumbai - 400005, INDIA}
\author{C. D. Bagdia}
\affiliation{Deptartment of Nuclear and Atomic Physics, 
Tata Institute of Fundamental Research, Mumbai - 400005, INDIA}
\author{L.C. Trivedi}
\affiliation{Deptartment of Nuclear and Atomic Physics, 
Tata Institute of Fundamental Research, Mumbai - 400005, INDIA}


\date{\today}

\begin{abstract}
In nuclear astrophysics, the experimental scenario is still unclear for BBN relevant energies from 30-300 keV. Here, we report our measurements of cross section and astrophysical S(E) factor for pd capture reaction using proton beam energies 66 keV, 116 keV and 166 keV. Realistic GEANT4 simulations were carried out in order to estimate the energy dependent efficiency of the detector. The measured cross section and astrophysical S(E) factor for three new incident proton energies are found to be in good agreement, within the errors, with the values reported in the literature.
\end{abstract}


\maketitle

\section {Introduction}
The radiative capture of proton on Deuteron  \textit{d(p,$\gamma$)$^3$He} is a reaction of great significance both for nuclear astrophysics and few-body nuclear physics.  From experimental standpoint the beam energy for this reaction is varied over a very wide range depending upon the exact nature of the problem one is interested in.  As far as the nucleosynthesis of $^3$He is concerned the relevant beam energy  varies from few keV to few hundred keV.  Broadly, one can consider three scenarios for the  production of $^3$He (or depletion of deuteron) from \textit{d(p,$\gamma$)$^3$He} reaction, namely, the Big Bang Nucleosyntheis (BBN), production in low-mass protostars, and production in low to medium mass stars like sun\cite{rolf,Stahler1988,Iocco2009,Smith1993}.
 In BBN the prevalent temperature at which the \textit{d(p,$\gamma$)$^3$He} reaction takes places is around 10$^9$K (T$_9$).  This temperature translates to a beam energy of around few hundred keV.  In contrast the nucleosynthesis of $^3$He from the capture of proton on deuteron inside stars (both stellar and protostellar)  happen at a much lower energy.  In protostars  the reaction takes place around 0.1 keV and in sun-like stars it is  around few keV. 
This particular reaction leads to the depletion of deuteron in all the environments, mentioned above with the reaction rates governing the D/H ratios\cite{Tytler1996}.  While the reaction rate for \textit{d(p,$\gamma$)$^3$He} is crucially important to understand the evolution of protostellar stars it is not all that significant in the case of stellar evolution.  This is primarily because of the fact that p-d capture follows a weak process of deuteron production in stellar environment whereas it is the very first process of nuclear ignition in a protostar.  \\
The LUNA experiment\cite{Casella2002} has provided high quality data for cross sections and astrophysical S-factors for the \textit{d(p,$\gamma$)$^3$He} reaction from 2.5 to 22 keV (CM), essentially covering the energy region of interest in stellar environment.  It is to be noted that the solar Gamow peak is around 9 keV for \textit{d(p,$\gamma$)$^3$He} reaction.   However, there is rather limited data for the BBN energy region from 30 to 300 keV.   It is quite evident from Ref. \cite{Marcucci2016}that the experimental data is very much in disagreement with the best polynomial fit for S-factor and thereby introduces the uncertainty in the cross section in the range of 6-10$\%$ level. Therefore, the measurements of cross section and S-factor for \textit{d(p,$\gamma$)$^3$He} reaction at BBN relevant energies are of great fundamental importance.\\

\begin{figure}[h]
\centering
\includegraphics[height=7.0cm, clip,width=0.5\textwidth]{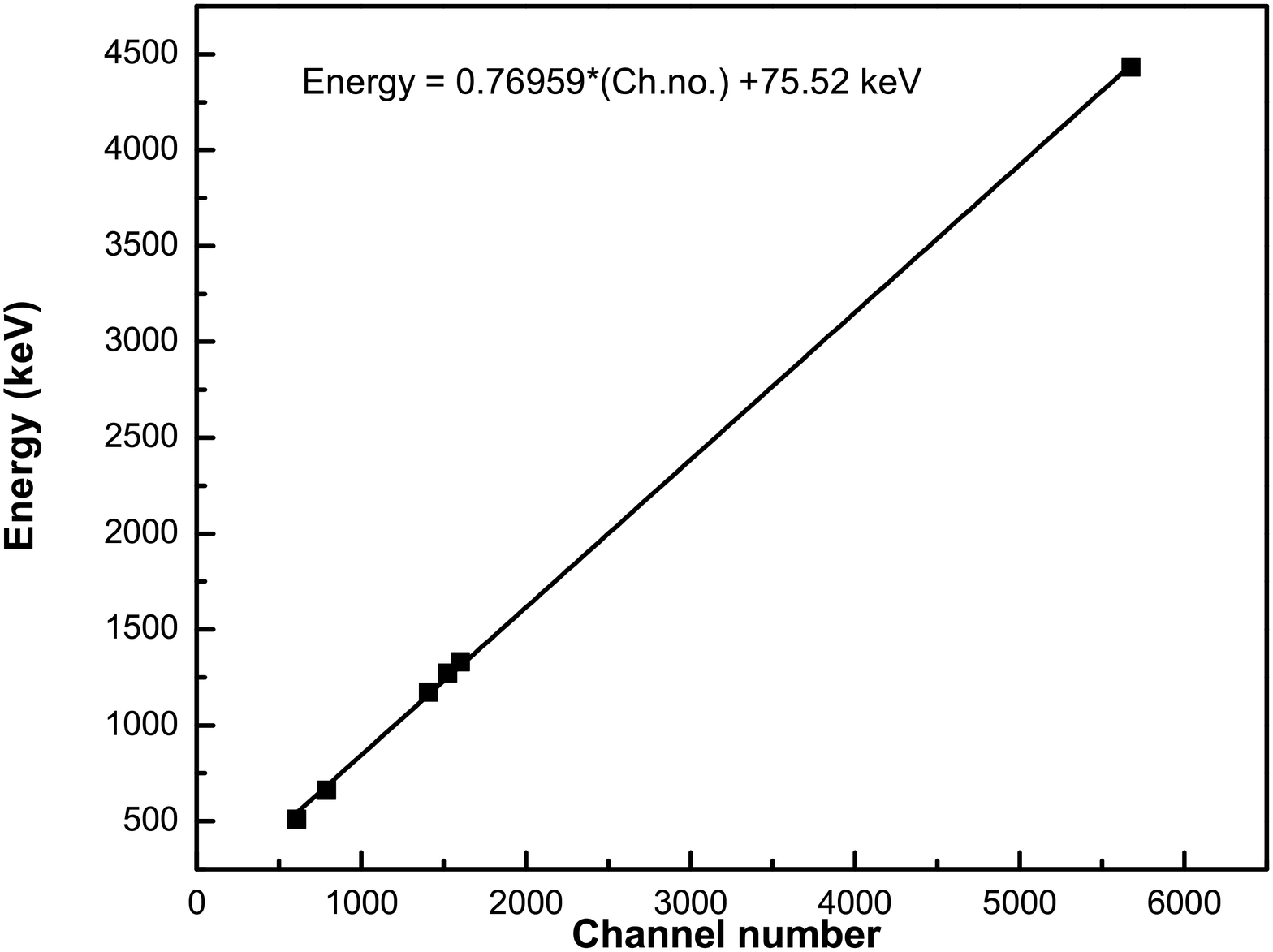}
\caption{\label{fig: 1} Linear responce of the large volume LaBr$_3$:Ce detector from 511 keV to 4.43 MeV.}
\end{figure}

There is also uncertainty about the BBN D/H\cite{Tytler1996} ratio from measurement of absorption of radiation from distant quasars in intervening clouds of gas.  It is therefore required to carry out more precision measurements yielding cross section and astrophysical S-factors in the 30 to 300 keV region.\\
On a different plane, the proton-deuteron (p-d) capture reaction is important for studying the bound $^3$He system\cite{Viviani1996,Fonseca1993}.  It is  an ideal laboratory to probe a variety of features, such as, Coulomb effect in a three-body system, role of meson exchange current\cite{Phillips1972} in nucleon-nucleon interaction, search for three-body force etc.  Considerable amount of experimental data and theoretical analysis exist in literature for this reaction with beam energies varying from few keV to few hundred MeV\cite{Casella2002,Griffiths1963,Bailey1970,Ma1997,Schmid1997,Bystritsky2008}.\\
This paper reports about our measurements of cross sections and astrophysical S-factors for the \textit{d(p,$\gamma$)$^3$He} reaction, ostensibly, in the BBN energy region.  Several experiments  have reported in past about proton-deuteron (p-d) capture cross sections at very low energy region\cite{Casella2002, Ma1997,Schmid1997,Bystritsky2008} below 80 keV.  In reference\cite{Schmid1995R} the authors have compared their results with that of \cite{Stahler1988}and have reported a deviation in S-factors of about 40 to 50$\%$ from previous results.  By far the most exhaustive set of data in the very low energy region of stellar synthesis ( 2 to 22 keV) have been reported by the LUNA collaboration\cite{Casella2002}.  In comparison there are much lesser number of data points available in the BBN region, namely, 30 to 300 keV.  Some of the oldest measurements were reported by Griffiths \textit {et.al.,}\cite{Griffiths1963} and Bailey \textit{et.al.,}\cite{Bailey1970}.  More recently Ma \textit{et al.,}\cite{Ma1997} have reported measurements for p-d capture in the range of ~ 70-200 keV.  Ma \textit{et.al.,}  reported a S(0) value 25$\%$ lower than what is presently used in astrophysical calculations.  Very recently, Marcucci \textit{et.al.,}\cite{Marcucci2016} have carried out detailed \textit{ab initio} calculations for the astrophysical S-factor for p-d capture in the energy region of BBN.  They have included both two- and three-nucleon interactions and one- and many-body contributions to the nuclear current operator.  Their predictions for the D/H ratio is found to be in excellent agreement with the experimental  value determined using the baryon density obtained from the Planck experiment.  They conclude that there is clearly a need for more experimental measurements of the cross sections and S-factors in the 30 to 300 keV region.\\
We have measured the cross section and S-factors for the \textit{d(p,$\gamma$)$^3$He} reaction at proton beam energies of 100, 175 and 250 keV.  The authors in references\cite{Ma1997} have measured the capture gamma rays using HPGe detectors surrounded by segmented NaI(Tl) anti-coincidence shield.  We have measured the gamma rays using a large volume Lanthanum Bromide (LaBr$_3$:Ce) detector.  This is, to the best of our knowledge, the first measurement of capture gamma rays for the \textit{d(p,$\gamma$)$^3$He} reaction using a large volume LaBr$_3$:Ce detector.  We discuss the experiment in the next section.  

\begin{figure}[h]
\centering
\includegraphics[height=7.0cm,width=0.5\textwidth]{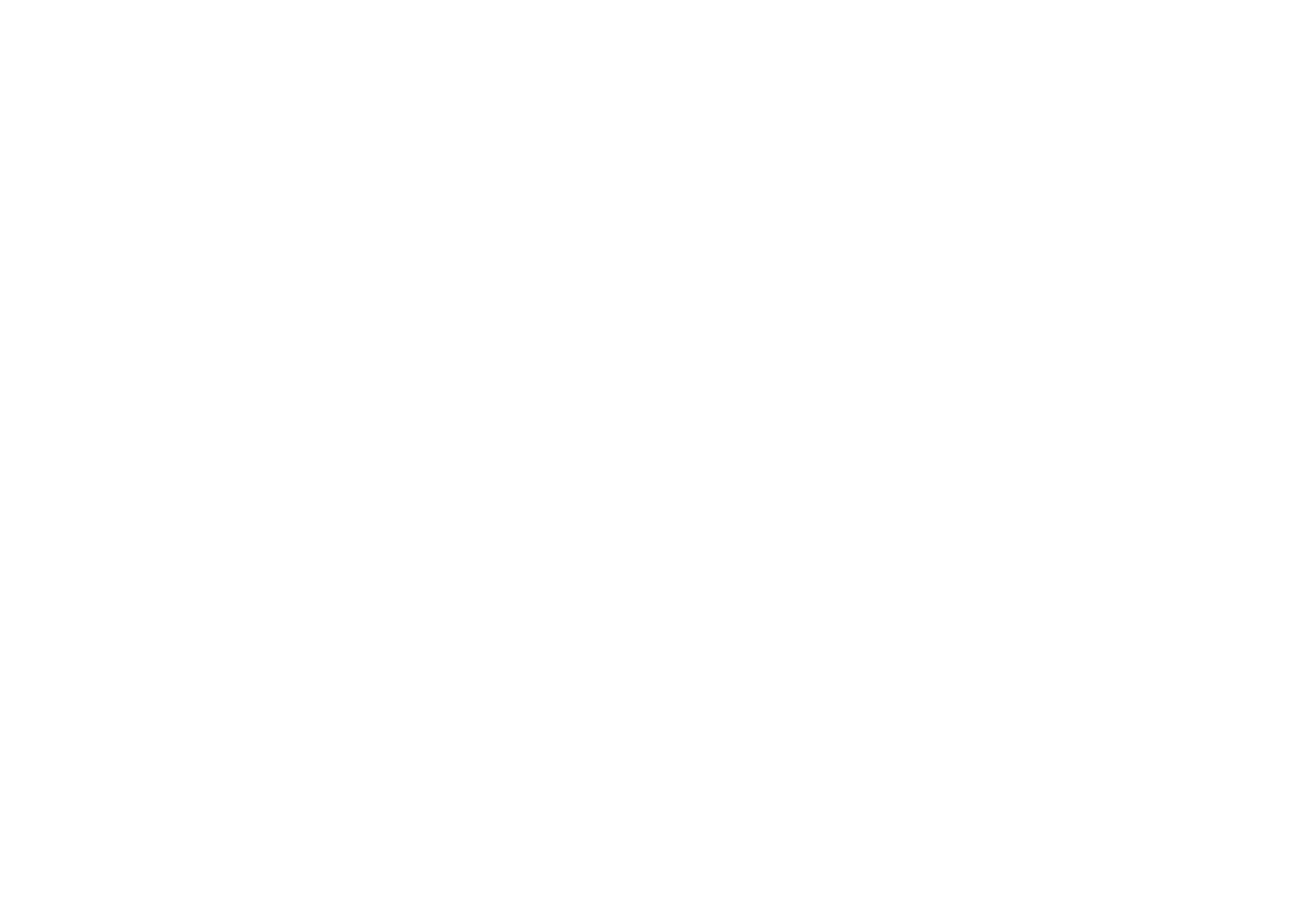}
\caption{\label{fig: 2} A typical spectrum for captured $\gamma$-rays at E=250keV.}
\end{figure}

\section{Experimental Details}
\label{sec:2}
The \textit{d(p,$\gamma$)$^3$He} reaction was carried out by bombarding a CD$_2$ target using proton beam of 100, 175 and 250 keV corresponding to CM energies of 66, 116 and 166 keV, respectively. The proton beam was obtained from the 14.5 GHz 300W Electron Cyclotron Resonance (ECR) ion source based low energy accelerator facility at TIFR, Mumbai\cite{ecr}.  The ion source is based on a Cu-plasma chamber.  The maximum extraction voltage of the ECR ion is 30 kV and the deck voltage can be raised up to 400 kV to accelerate the ions further.  The charge states of ions are analyzed using a 90$^\circ$ bending dipole magnet.  The ECR ion source can deliver a steady high current (few tens of $\mu$A) for long duration during the experiment even at minimum extraction voltage.  The deuterated polyethylene (CD$_2$) target had thickness of 5.1$\times$10$^{17}$ atoms/cm$^{2}$.  The $\gamma$-rays produced from the capture of proton on deuteron were detected using a highly efficient large volume cylindrical LaBr$_3$:Ce scintillation detector. The detector is 6$\,{''}$ long with 3.5$\,{''}$ diameter and is surrounded by Al casing of 0.8mm thickness. The detector was placed close to the target chamber at 4.5 cm from the centre of the target.  The energy calibration was carried out using standard low energy gamma sources, namely, $^{137}$Cs (661.6 keV), $^{60}$Co (1173, 1332 keV), $^{22}$Na (511, 1274 keV).  The gamma-rays of interest from \textit{d(p,$\gamma$)$^3$He} reaction for the beam energies of 100,175 and 250 keV vary from 5.55 to 5.71  MeV.  Therefore, an Am-Be source producing 4.433 MeV high gamma-rays was also used  for calibration, and measurement of energy resolution and linearity up to 4.43 MeV. Fig.1 shows the very good linear responce of the detector from 511 keV to 4.43 MeV.\\

\begin{figure}[h]
\centering
\includegraphics[height=7.0cm, clip,width=0.5\textwidth]{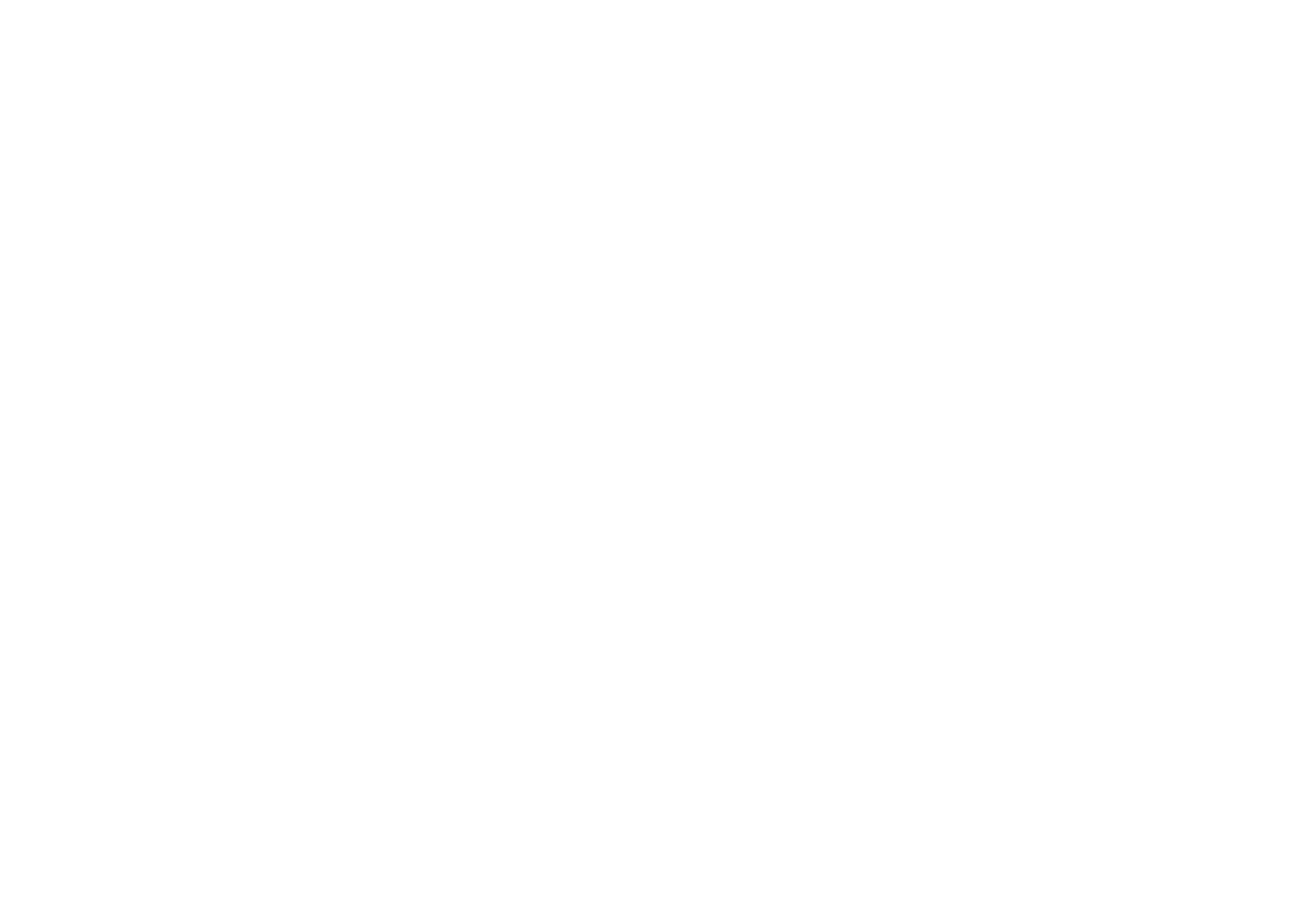}
\caption{\label{fig: 3} (Colour online) A typical spectrum for captured $\gamma$-rays at E=175keV.}
\end{figure}

The salient feature of the current experiment is the use of a large volume, cylindrical ( 3.5\,{''}$\times$6\,{''} )  LaBr$_3$:Ce detector. The significantly superior properties of the LaBr$_3$:Ce crystals over other scintillators are responsible for growing application of these crystals, namely,  in nuclear spectroscopy \cite{Quarati2011}, astronomical measurements\cite{Gostojic2015,
Gostojic2016}, medical imaging\cite{Pani2015,Kuhn2004}, geological applications \cite{Iltis2006,bizarri2006} etc. The properties of the detector, used in the present work, such as linearity, resolution (energy and time), detection efficiency for gamma rays in the energy range of few hundred keV to 22.5 MeV have been measured by the authors and can be found in Ref.\cite{Mazumdar2013}.

 The superior energy and timing resolution and higher efficiency of detection of LaBr$_3$:Ce over NaI(Tl) make it distinctly advantageous to use in experiments like the current one.  While the usage of large volume LaBr$_3$:Ce has picked up in recent years\cite{Mazumdar2014, Camera2014}, no experimental measurements, to the best of our knowledge, has so far been reported using large volume LaBr$_3$:Ce detectors in reactions of nuclear astrophysics.  The detector was operated at -600V to provide a perfectly linear response up to the energy of interest.  The energy and timing signals were drawn from the second last dynode and the anode respectively.  The energy signal was fed to a charge sensitive pre-amplifier and the  output was fed to a spectroscopy amplifier.  The shaped and amplified energy signal was recorded in a peak sensing ADC and data were collected in events mode.  The zero-crossover technique was used to generate the pileup spectrum.  The beam was stopped on a Faraday cup beyond the target and was measured using a beam current integrator.   The large volume detector was shielded from low energy background radiation by 4$\,{''}$ of lead shielding. \\
 
\section{Measurement and Analysis Method}
\label{sec:3} 

\begin{figure}[h]
\centering
\includegraphics[height=7.cm, clip,width=0.5\textwidth]{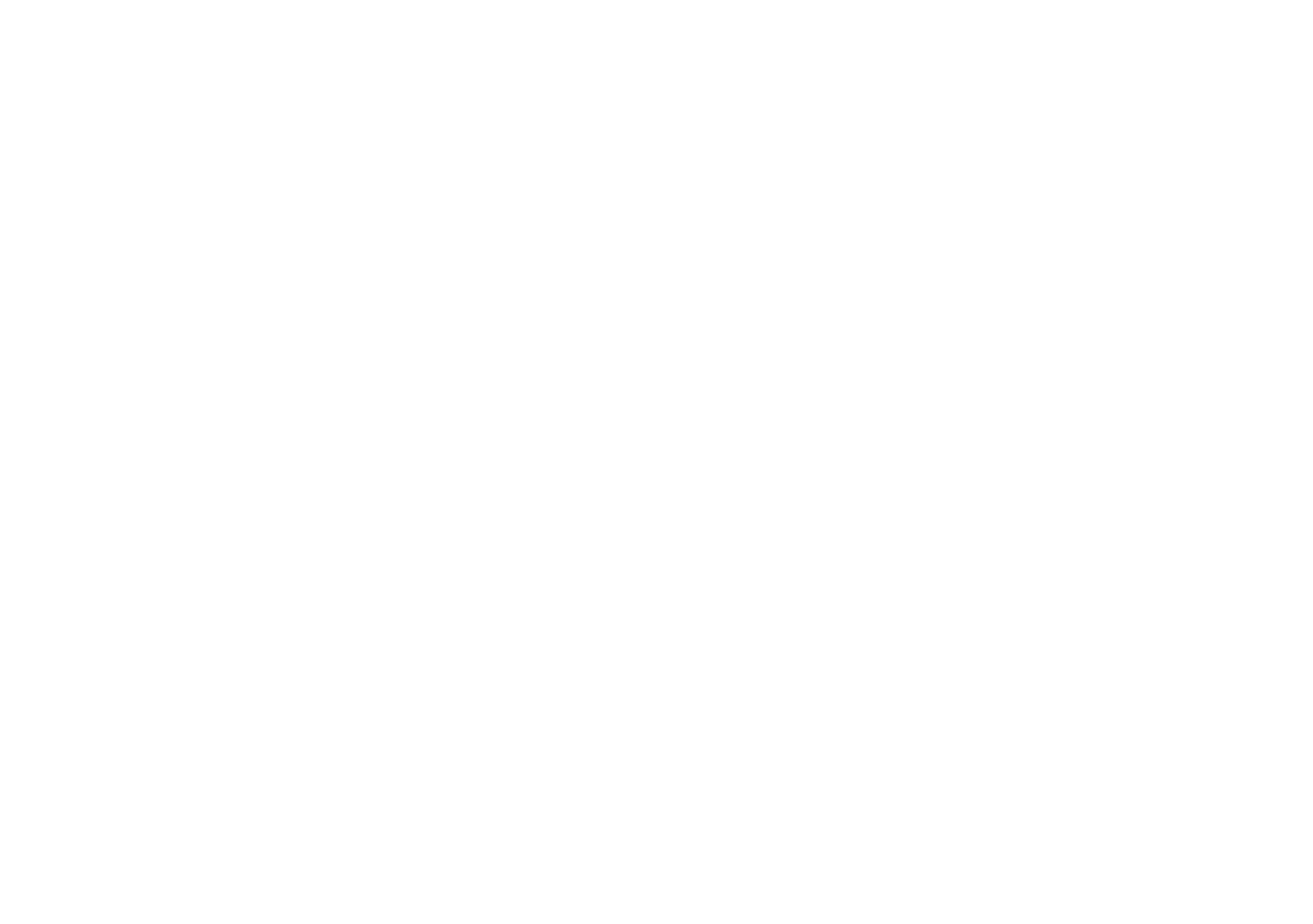}
\caption{\label{fig: 4} (Colour online) A typical spectrum for captured $\gamma$-rays at E=100keV.}
\end{figure}

The cross section and astrophysical S-factor were measured at three different beam energies, namely 100, 175, 250 keV in lab system.  Figure 2,3 and 4 present the measured gamma ray spectra  at the three beam energies.  The very good energy resolution of LaBr$_3$:Ce allows  distinct separation of the photo-peak and the first escape peak.  The cross section and astrophysical S factors\cite{rolf} are defined as
\begin{equation}
\sigma_{cm} = \frac{Yield}{n_{1}n_{2}(\epsilon{_1} \epsilon{_g})}
\end{equation}

where 
n${_1}$, n${_2}$, $\epsilon{_1},  \epsilon{_g}$
denote the total number of incident particles on the target, the target thickness, the intrinsic photo-peak efficiency and the geometric efficiency and

\begin{equation}
S(E)_{cm} = \frac{E_{cm}\sigma(E)_{cm}}{e^{-2\pi\eta}}
\end{equation}
\\
where $\eta$ denotes the Somerfield parameter and 2$\pi\eta$=31.29Z$_1$Z$_2$($\mu/E$)$^{1/2}$, $\mu$ and $E$ are the reduced mass of the system in amu and energy in cm
system in keV. 
\\

\begin{figure}[h]
\centering
\includegraphics[height=8.cm, clip,width=0.5\textwidth]{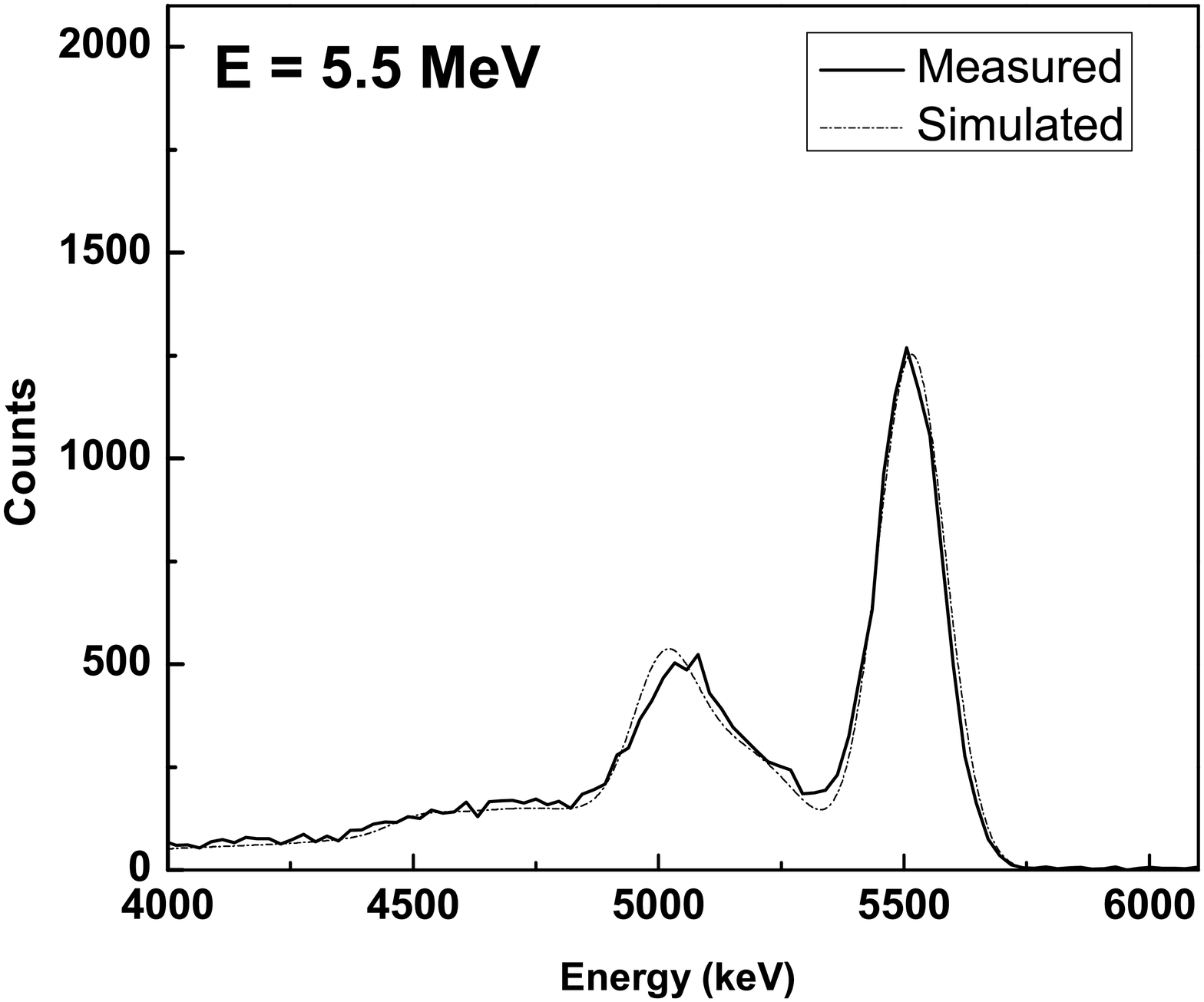}
\caption{\label{fig: 5} (Colour online) Measured 5.5 MeV $\gamma$-ray responce along with GEANT4 simulation.}
\end{figure}

Background spectra were recorded with beam through a blank frame for the same duration of time after each experimental run.  The back ground spectra were subtracted from the in-beam spectra at all the three energies. The capture of proton by $^{12}$C in the target can generate gamma-rays of energy around 2 MeV.  This is because of the Q value of the reaction being 1.943 MeV.  This being sufficiently away from the gamma peak due to proton capture on deuteron a rather high energy threshold of 2 MeV was used in the experiment.  The extraction of the production cross section of gamma rays crucially depend upon the use of correct detection efficiency.  The detection efficiencies of the detector used in this work have been measured for a large number of gamma ray energies from 662 keV ($^{137}$Cs line) to 30 MeV.  The measurements have been carried out using calibrated gamma ray sources, proton induced nuclear reactions and high energy monochromatic gamma rays from  the Duke Free Electron Laser (FEL) facility.  In addition, we have simulated the response of the detector using realistic GEANT4 simulations\cite{geant}.\\

\begin{figure}[h]
\centering
\includegraphics[height=7.cm, clip,width=0.5\textwidth]{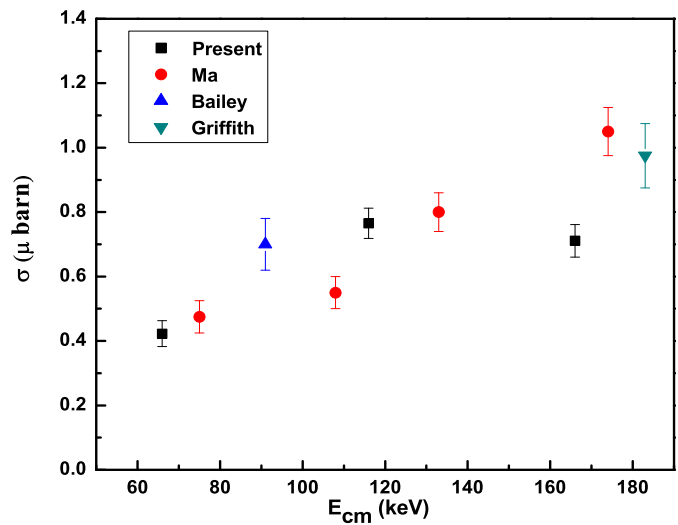}
\caption{\label{fig: 6} (Colour online) Measured cross-section values as function of Energy is shown together with existing experimental data \cite{Griffiths1963,Bailey1970,Ma1997}.}
\end{figure}

\section{Simulation}
\label{sec:4}

The detection efficiency ( photopeak or total detection ) is the product of the intrinsic and geometric efficiencies.  In a realistic simulation in addition to the geometric factors one needs to consider the environment surrounding the detector and all the absorbing materials between the  target and the face of the detector.  In our simulations the environment, namely the wooden support structure for the detector and all the absorbing materials in front of it were considered.  The Low Energy Electro-Magnetic Physics package\cite{Chauvie} was used in the physics list class. The radioactive Decay Module has been used to simulate the decay of the point sources, namely, $^{137}$Cs (661.6 keV), $^{60}$Co(1173, 1332 keV), $^{22}$Na(511, 1274 keV), Am-Be(4.438 MeV) and 5.5 MeV $\gamma$-rays\cite{hauf}.  The energies and branching ratios of the $\gamma$-sources used in the simulation were taken from NNDC data table\cite{nndc}. The simulation was started by generating an isotropic emission of $\gamma$-rays by assuming a point source. The total sample of 10$^6$ $\gamma$-ray events were generated and distributed over the detector surface.\\

\begin{figure}[h]
\centering
\includegraphics[height=7.cm, clip,width=0.5\textwidth]{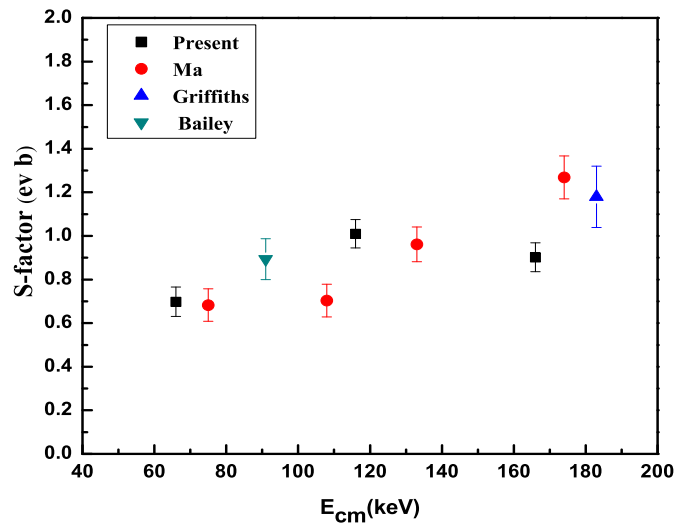}
\caption{\label{fig: 7} (Colour online)Measured astrophysical S-factor values as function of Energy is plotted along with existing experimental data \cite{Griffiths1963,Bailey1970,Ma1997}.}
\end{figure}

\begin{table}
\centering
\caption{Photo-peak and total detection efficiencies of large volume LaBr$_3$:Ce
detector calculated by GEANT4 simulation.}
\vspace{5mm}
\begin{tabular}{|c|c|c|c|c|}
\hline 
Energy&Photo-peak&Total detection \\
(keV)&($\%$)&($\%$)\\
\hline
 662& 46.1$\pm$0.95 & 71.1 $\pm$1.24\\
1000 & 38.8$\pm$0.92 & 69.2$\pm$1.32 \\
4433 & 16.8$\pm$0.83 & 61.4$\pm$1.41 \\
5000 & 14.8$\pm$0.95 & 56.3$\pm$1.09 \\
5500 & 13.6$\pm$0.87 & 52.5$\pm$1.12 \\
\hline
\end{tabular}
\label{table1}
\end{table}
Fig.5 shows the measured gamma ray spectrum for 5.5 MeV monochromatic photons along with the GEANT4 reproduction.  We also show in Table I some of the values of photopeak efficiencies as calculated by GEANT4. The energy dependent resolution was fed in to the GEANT4 calculation to generate the gamma ray spectral profiles.  The dotted lines in Fig.2,3 and 4 present the GEANT4 simulations of the detector response to the measured gamma rays  at three different beam energies.  The very good reproduction of the profiles provide the confidence to use the simulated values of efficiency to extract the cross sections.  The cross sections and astrophysical S factors are provided in the table 2.

\begin{figure}[h]
\centering
\includegraphics[height=7.cm, clip,width=0.5\textwidth]{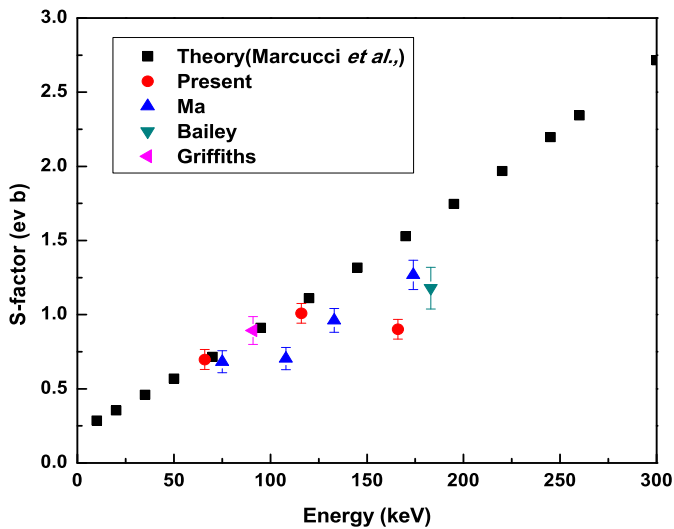}
\caption{\label{fig: 8} (Colour online)Measured astrophysical S-factor values and the reported literature \cite{Griffiths1963,Bailey1970,Ma1997} are shown along with calculated values by Marcucci \textit{et al.,}\cite{Marcucci2016}.}
\end{figure}

\section{Results and discussions}
\label{sec:5}
\begin{table}
\centering
\caption{The measured cross sections and S-factors.}
\vspace{5mm}
\begin{tabular}{|c|c|c|}
\hline 
$E_{cm}$&$\sigma$& S-factor \\
(keV)&($\mu$b)&(ev b)\\
\hline
166.66 & 0.71$\pm$0.0504 & 0.902 $\pm$0.066\\
116.66 & 0.73$\pm$0.0406 & 0.967$\pm$0.064 \\
66.66 & 0.42$\pm$0.040 & 0.698$\pm$0.067 \\
\hline
\end{tabular}
\label{table1}
\end{table} 
 
We have extracted the cross section and astrophysical S factors at three new energies in the region important to understand the production of $^3$He from proton capture of deuteron during Big Bang Nucleosynthesis (BBN).  Fig.6 and 7. present our measured three new cross sections and S-factors for 66,116,166 keV along with the other existing data from previous workers. 
Our measured values for both $\sigma$ and S factor are in good agreement of the global variation with energy.  We have also compared our measured values with the recent calculations of Marcucci \textit{et al}.,\cite{Marcucci2016}.
  In Fig.8. measured values fall within the band predicted by Marcucci \textit{et al}.,\cite{Marcucci2016}.  The importance of the precision measurements of  \textit{d(p,$\gamma$)$^3$He} cross sections have been discussed in the introduction.  The very encouraging results of our present measurements provide sufficient impetus and confidence for future measurements covering   much larger beam energies from 30 to 300 keV.  These measurements are currently underway and will be reported in due course.  

\section*{Acknowledgment}
One of the authors (MD) acknowledges the financial support from the Ministry of Human Resource Development, Government of India, and the author (G. Anil Kumar) acknowledges the partial financial support received from DST, Government of India as part of the fast track project (No: SR/FTP/PS-032/2011).




\end{document}